\documentclass[3p]{elsarticle}
\usepackage{lineno,hyperref}
\usepackage{color,graphicx}
\usepackage{url}
\usepackage{amsmath}
\usepackage{algorithmic}
\usepackage{algorithm}
\usepackage{epsfig}
\usepackage{caption}
\usepackage[T1]{fontenc}
\usepackage{txfonts}
\usepackage{array,multirow,graphicx}

\modulolinenumbers[5]
\begin{document}
\begin{frontmatter}
%A Review of Milestone Mechanisms in Small Cell Planning\\
%All That Matters in Small Cell Planning: A Review\\
\title{Evolution of Small Cell from 4G to 6G: Past, Present, and Future}
\author{Vanlin Sathya}
\address{Department of Computer Science, University of Chicago, Illinois, USA.}
\fntext[myfootnote]{Email: vanlin@uchicago.edu (Vanlin Sathya)}

\begin{abstract}
{
To boost the cellular system's capacity, the operator's have started to reuse the same licensed spectrum by deploying 4G LTE small cells (\emph{i.e.,} Femto Cells) in the past. But in time, these small cell licensed spectrum is not sufficient to satisfy future applications like augmented reality (AR) and virtual reality (VR). Hence, cellular operators look for alternate unlicensed spectrum in Wi-Fi 5 GHz band, later 3GPP named as LTE Licensed Assisted Access (LAA). The recent and current roll-out of LAA deployments (in developed nations like the US) provides an opportunity to understand coexistence's profound ground truth. This paper discusses a high-level overview of my past, present, and future research works in the direction of small cell benefits. In the future, we shift the focus onto the latest unlicensed band:  6 GHz, where the latest Wi-Fi version, 802.11ax, will coexist with the latest cellular technology, 5G New Radio (NR) in unlicensed.
\begin{keyword}
Femtocells, Small cells, LAA, NR-U, Wi-Fi
\end{keyword}
}
\end{abstract}
\end{frontmatter}

\section{Introduction}
The growing penetration of high-end consumer devices (smartphones, tablets, etc.) running bandwidth-hungry applications (e.g., mobile multimedia streaming) has led to a commensurate surge in demand for mobile data (pegged to soar up to 77 exabytes by 2022). An anticipated second wave will result from the emerging Augmented/Virtual Reality (AR/VR) industry and, more broadly, the Internet-of-Things that will connect an unprecedented number of intelligent devices to next-generation (5G and beyond) mobile networks. These must, therefore, greatly expand their aggregate {\em network} capacity to meet this challenge. It is achieved by combining approaches, including multi-input, multi-output (MIMO) techniques, network densification (\emph{i.e.,} deploying small cells), and more efficient traffic management and radio resource allocation. 

On the other side cheap, fast, and portable computing devices with ubiquitous wireless connectivity~\cite{giluka2014class,dama2016novel,dama2016feasible} can revolutionize the personal computing~\cite{kala2019designing} landscape by creating an opportunity to design an unprecedented array of new services and applications. Keeping the same philosophy in mind, we have focused mainly on maximizing the experienced data rate of an end-user by offering increased bandwidth through the coexistence of licensed wireless services (e.g., Long Term Evolution (LTE)) an unlicensed band~\cite{E1,sagari12, LWS22, LWS11}. To achieve the goal of throughput maximization~\cite{Cisco}, we focused on minimizing the interference and handling frequent handovers. It is done by optimally placing the small cells~\cite{ramamurthy2019dynamic,kala2018exploring} (\emph{i.e.,} a miniature base station, specifically designed to extend the data capacity, speed, and efficiency of a cellular network~\cite{lokhandwalaeai}) and controlling their emitting power in a dense small cell deployment scenario.

{\hskip 2em} 5G LTE, which operates in the licensed band and 802.11 wireless LAN (Wi-Fi)~\cite{kala2019socio,kala2018icalm}, which operates in an unlicensed band, has some fundamental structural differences. For example, LTE control is where a base station (BS) exclusively allocates the radio resources to the users, interference due to concurrent transmission by the users handled. On the other hand, Wi-Fi~\cite{ieee, W4} follows a distributed approach~\cite{boe,bianchi} where each user independently contests to occupy the channel, thereby concurrent transmission results in interference. The main motive of LTE/Wi-Fi coexistence in unlicensed bands for LTE users in case of very few or no Wi-Fi users. So, the research on the fair coexistence of LTE/Wi-Fi mainly focuses on the intelligent use of the unlicensed band by the LTE users to keep the Wi-Fi users unaffected so that the aim of formation of the unlicensed band remains unaltered. 

{\hskip 2em}  The standard development community has accepted two mechanisms for using the unlicensed band by LTE. These mechanisms are licensed assisted access (LAA) and LTE Unlicensed (LTE-U). LAA follows the same approach of sensing the unlicensed channel, called Listen Before Talk (LBT), as Wi-Fi. LTE-U estimates its duty cycle to access the unlicensed channel based on the various parameters such as interference, type of traffic, and load on the track.  In our recent research, we have focused on both LAA and LTE-U mechanisms. In both mechanisms, we keenly observe the various aspects of real-time, which can adversely affect the Wi-Fi users and were ignored by the existing literature. Some of the observations are a) static channel allocation to LTE-U node in an unlicensed band, b) difficulty in association to the Wi-Fi access point (AP) by the Wi-Fi users due to high duty cycle (\emph{i.e.,} repeating ON and OFF intervals in the medium) in LTE-U, c) a considerable reduction in the duration of duty cycle in LTE-U if several surrounded APs considered in its estimation. We used a Machine Learning approach to propose solutions based on these observations. As part of my future research plan, We want to explore the research challenges in the fair coexistence of LTE and Wi-Fi on the 6 GHz band used as an unlicensed band. Apart from that, We plan to provide Machine Learning (ML) based solutions to some of the existing problems on LTE/Wi-Fi coexistence on 5 GHz. These problems are an efficient use of high bandwidth by Wi-Fi users and optimal channel selection by both LAA BS and a Wi-Fi AP in a multi LTE operators-multi AP scenario. ML algorithms are used to closely observe the system's behavior on different conditions to make intelligent decisions.\\

The paper is organized as follows. Section 2 provides a brief overview of 4G and 5G Heterogeneous Networks. Section 3 describes the associated challenges and solutions for the past and present small cell deployments. Section 4 focuses on the future and recent NR-U small cell in 6 GHz. Finally, conclusions and future research directions are presented in Section 5.\\

\section{4G \& 5G Heterogeneous Networks (HetNets)~\cite{krishna2014dynamic,madhuri2014dynamic}:} 
\begin{figure}	
\begin{center}
			\includegraphics[width=12.5cm]{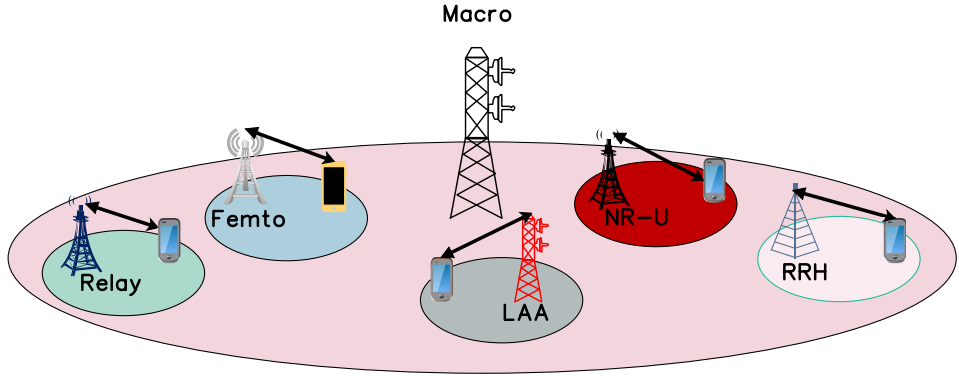}
			\caption{4G and 5G Heterogeneous Network}
			\label{hetnet}
			\end{center}
			\end{figure}

Presently, cellular network users are not only those which generate mostly downlink traffic (\emph{i.e.,} web browsing, downloading) but also combination of users generating symmetric (both uplink (UL) and downlink (DL)) traffic (\emph{i.e.,} social networking, gaming) and users generating uplink traffic (\emph{i.e.,} M2M/IoT). In order to provide better connectivity and high data rates to these users, low power network nodes, called as small cells, are being deployed. Presence of such diverse traffic generating users and small cells with different transmit powers and sizes, has turned cellular networks from homogeneous to heterogeneous in nature. A heterogeneous network (HetNet) consists of a Macro cell augmented with various types of small cells to address the challenge of enhancing system capacity and coverage. Examples of small cells are micro cell, pico cell, relay, Remote Radio Head (RRH), Femto, Licensed Assisted Access (LAA), New Radio in Unlicensed (NR-U), and Femto cell~\cite{chaganti2013efficient,sathya2013enhanced} as shown in Fig.~\ref{hetnet}.  Table~\ref{hetnettable} shows characteristics of 
various types of cells. The description of various types of small cells are as follows.
\begin{table}[htb!]
\caption{Characteristics of heterogeneous cells in 4G and 5G}
\centering	
\begin{tabular}{|p{3.2cm}| p{2cm}|p{2cm}|p{4cm}| p{2cm}|}
\hline\hline
\bfseries
\ \hspace{0.7cm}Technology &\bfseries  Placement &\bfseries Transmit Power  &\bfseries Backhaul Characteristic &\bfseries Number of Users\\
\hline	
Macro BS& Outdoor & 46 dBm & Dedicated wireline & 1000-2000 \\
\hline
Pico or Micro cell &Outdoor & 30 dBm & Dedicated wireline & 100-200 \\
\hline
RRH &Outdoor or Indoor &30-35 dBm & Dedicated wireline & 100-200 \\
\hline
Relay &Outdoor or Indoor & 30-35 dBm & Wireless out-of-band or in-band & 60-100 \\
\hline
Femotcell &Indoor & 20-23 dBm & Residential or enterprise broadband & 10-30 \\
\hline
LAA~\cite{WL,7} &Outdoor or Indoor & 20-23 dBm & Residential or enterprise broadband & 10-30 \\
\hline
NR-U &Outdoor or Indoor & 20-23 dBm & Residential or enterprise broadband & 10-30 \\
\hline
\end{tabular}
\label{hetnettable}
\end{table}

\begin{itemize}
\item \textbf{Pico or Micro BS:} These are deployed in an outdoor environment to cover a radius up to 300 m. The pico BS (\emph{i.e.,} 30 dBm) is smaller than that of Macro BS. Pico BS has a dedicated $X3$ backhaul connection to the Macro BS for coordination.
\item \textbf{Relay:} Relay BS acts as a repeater. It receives the data/signal from the Macro BS and transmits/boosts the data to the relay connected users. The link between the relay and Macro BS is also wireless. Normally the relay BS(s) is the preferred choice of operators to extend the coverage region (e.g., Hilly region, rural areas) of the Macro BS.
\item \textbf{RRH:} Unlike traditional BS, RRH is a radio transceiver component that performs only the transmission and reception of In-phase Quadrature (IQ) samples. The remaining BS processing is done at a centralized cloud data center by Baseband Unit (BBU) pool. The BS's cost will come down by performing centralized processing, which directly reduces CAPEX and OPEX. 
\item \textbf{LAA:} Unlike traditional Femtocells, LAA operates on the 5 GHz unlicensed spectrum, which does implement an LBT protocol similar to that used by Wi-Fi, with different values for parameters such as sensing threshold and transmission intervals.
\item \textbf{NR-U:} Unlike LAA cells, NR-U operates on the 6 GHz unlicensed spectrum, which implements an LBT protocol to fair share the spectrum protect the incumbent users. 
\end{itemize} 
The advantage of HetNets is as follows.
\begin{enumerate}
 \item \textit{cell range expansion (CRE)~\cite{ramamurthy2016improving,kumar2017enhancing,akilesh2016novel}:} Increasing or decreasing the transmit power (or coverage region) based on the load in small cells will boost the overall system throughput.
 \item \textit{Integrating Macro and small cells:} Improving user throughput by dual connectivity~\cite{martolia2017enhancing} (\emph{i.e.,} Macro and one small cell).
 \item \textit{Self-organizing network (SON):} In HetNets, small cells are deployed in huge number. The operator can provide each small cell with SON features, which aim to automatically configure and optimize the network, reducing the human effort. It plays a key role in improving OAM (operation, administration, management). 
\end{enumerate}

\subsection{LTE Femtocell Networks}\label{sec:LTEFemNet}
\par The existing Macro Base Stations (MBS) cannot satisfy mobile users because of most users' huge data demand and indoor locality. Reports by Cisco tell that 70\% of the traffic is generated in indoor environments such as homes, enterprise buildings, and hotspots. Hence, mobile operators must improve coverage and capacity`\cite{sathya2013dynamic} of indoor environments. But the basic problem with the existing MBS (or small outdoor cells with shorter coverage) is that they can only boost data rates of Outdoor User pieces of equipment ($OUEs$). But, they cannot do the same for Indoor User pieces of equipment ($IUEs$) because it is difficult for electromagnetic signals to penetrate through walls and floors. Owing to numerous obstacles in the communication path between MBS and $IUEs$ inside the building, radio signals attenuate faster with an increase in the distance. Thus, $IUEs$ receive low signal strength (\emph{i.e.,} Signal-to-Noise Ratio, SNR) compared to outdoor users. Hence, mobile operators must improve coverage and capacity in indoor environments.

 \par As a solution, Femtocells are being deployed by both operators and end customers. Femtocell is a low-cost, low-power consuming cellular base station that operates only in a licensed spectrum and designed for outdoor and indoor communication. The range of Femtocell is 100-150 meters for enterprise environments consuming 100 mW power. A home-based Femto (HeNB) can serve 4-5 users, whereas an office-based Femto can serve a maximum of 64 users. Each Femto requires a backhaul connection to the evolved packet core (EPC). Advantages of using Femtos are described as follows: \\ \\
\noindent \textbf{Operator Advantages :}
\begin{enumerate}

\item The operator can increase the network capacity.
\item The operator can reduce Operational expenditure (OPEX) and Capitational expenditure (CAPEX).
\item The operator can reduce Backhaul cost.
\item The operator can reduce Traffic overload on MBSs.
\end{enumerate}
\noindent \textbf{User Advantages :}
\begin{enumerate}
\item Improved Quality of Experience (QoE).
\item Improved energy efficiency/battery life.
\end{enumerate}

\subsection{Architecture of Indoor LTE Femto cells}\label{sec:LTEindoorFem}
In the LTE HetNet system's architecture, Femtos are deployed inside the building and connected to a Femto Gateway (F-GW) over the S1 interface. F-GW is mainly used to reduce the load on MME. It acts as a virtual core network to Femtos. The F-GW gets assigned with an eNB ID, and thus F-GW is considered another eNB by the MME. The X2 interface~\cite{X1,X2,X3,X4,X5} is introduced between Femtos of enterprise Femtocell networks to avoid inter-cell interference and directly route the data and signaling messages among Femtos, thereby reducing the load on LTE core network and offering better coordination among Femtos.   
\subsection{Access Modes in Femto}
Since Femtos~\cite{vanlin2013dynamic} are deployed for offering high data rates to indoor (paid) users in enterprise and residential buildings, each Femto is configured with a list of subscribers called Subscriber Group (SG) such that only the users in the SG can access the Femto. The users not belonging to this list are called Non-SG (NSG), and they may not be served by the Femto even when they are close to the Femto. Following access modes are defined for Femtos:
\begin{itemize}
\item \textbf{Open access:} The open-access mode allows all users (\emph{i.e.,} SG \& NSG) to access the Femto without any restriction.
\item \textbf{Closed access :} The fast-access mode permits only authorized users (\emph{i.e.,} SG) to access the Femto.
\item \textbf{Hybrid access:} The hybrid access~\cite{ghosh2017novel} is the combination of both open and closed access. It allows all users (\emph{i.e., SG \& NSG}) by providing preferential access for SG users over NSG users.
\end{itemize}

\subsection{LTE Licensed Assisted Access}
3GPP specifies LTE-LAA in Release 13 adopted the LBT approach for coexistence with Wi-Fi and supported only DL transmissions in the unlicensed band:  a secondary cell (sCell) aggregated with a licensed primary cell (pCell). Enhanced LAA (eLAA), as specified in Release 14, supports UL operation in the unlicensed band. However, the legacy LTE UL schedule continued to be used in eLAA, thus increasing the processing delay in scheduling grants due to LBT procedures. Hence, in April 2017, 3GPP started the "further eLAA" (FeLAA) working group (in Release 15) to improve LAA DL and UL performance through enhanced support for autonomous UL transmissions.  In the proposed FeLAA, a UL transmission ought to receive a grant from the eNB before the transmission, which solves the constraint imposed by the legacy eLAA. Most of these features have not been tested in the field before deployment. As LTE-LAA  deployments are being rolled out in major cities in the US, they offer an opportunity for real-world testing.

\subsection{5G Small Cell in Unlicensed Deployment}
The coexistence of small-cell LTE-U~\cite{A1, BI, A4, A3, 8, PL, cts, blank} and Wi-Fi networks in unlicensed bands at 5 GHz is a topic of active interest~\cite{baswade2016unlicensed,iqbal2017impact}, primarily driven by industry groups affiliated with the two (cellular and Wi-Fi) segments. In contrast, there is a body of analytical work~\cite{info, W2, W3, 3} exploring the coexistence of LTE-U and Wi-Fi, our focus in this project has been on real-time measurements ~\cite{xu2020understanding,narayanan2020lumos5g} and real-time deployment aspects of such coexisting networks. Coexistence is a topic that has seen little traction in the existing literature. As per the scope of this project, we actively design, analyze, and implement wireless network algorithms in simulation (using ns-3), in a real-time National Instrument (NI) unlicensed coexistence test-bed~\cite{mehrnoush2018analytical,mehrnoush2018fairness} and also conduct measurements and analysis on recently deployed LAA in the Chicago area.

In our previous work~\cite{mehrnoush2018analytical,mehrnoush2018fairness}, we investigated various aspects of coexistence between the two principle variants of LTE in the unlicensed bands, LTE-U and LAA,  and Wi-Fi~\cite{wilhelmi2019potential}. For LTE-U, we analyzed the effect of the LTE-U duty cycle~\cite{N1,E2,E3,3GPP} on the performance of the association of Wi-Fi. We demonstrated~\cite{sathya2019auto,adam2019detection,sathya2020machine,dziedzic2020machine} that using a high duty cycle adversely impacted Wi-Fi's ability to access the channel due to the Wi-Fi beacon transmission process's disruption. Therefore, we recommended using a lower duty cycle even if there was no Wi-Fi present in the channel to enable fair access to a new Wi-Fi access point that may wish to use the channel. We also developed machine-learning-based algorithms to determine the number of Wi-Fi APs on the air to enable an appropriate duty cycle setting. However, since LTE-U is not considered for wide deployment by industry, we switched our attention to evaluating LAA coexistence.

We have made tremendous progress toward understanding Wi-Fi and LAA's coexistence behavior in our previous work's unlicensed bands. Our theoretical analysis, corroborated by detailed system-level simulations using ns-3 and use of software-defined-radios, demonstrates that coexistence is improved substantially when the two systems treat each other symmetrically. That is when Wi-Fi and LAA use the same detection threshold to defer to each other. In our recent work~\cite{sathya2020measurement}, we added a new dimension by performing detailed measurements of deployed LAA networks by the three major carriers, Verizon, AT\&T, and T-Mobile, in Chicago. We conducted these measurements using off-the-shelf and custom-designed apps to extract detailed network information via APIs on Android smartphones. We believe this to be the first such exercise in academia and the measurements revealed several interesting new directions, which we will continue to research. Two such topics are: (i) Though most LAA deployments are outdoors and Wi-Fi's are indoors, the client devices that connect to these networks can be used outdoor/indoor; this results in hidden-node scenarios worse by the fact that two systems do not decode each other's signals (ii) Most academic analyses have focused on coexistence in a single 20 MHz channel, Whereas our measurements reveal that LAA usually aggregates three unlicensed channels, therefore increasing Wi-Fi's impact. These results have been presented to the industry as well and been incorporated into recommendations by Cisco.

 \section{Associated Challenges and Existing Solutions for past and present small cell deployments}
 %include related work in here
In this section, we discuss various issues and challenges while deploying small cells in LTE HetNets. Also, we discuss existing solutions to address these issues and challenges. All the challenges boil down to finding a solution to improve the system performance.
Some of the important challenges are discussed below:\\
\begin{enumerate}
\item Small Cell Placement in 4G LTE\\ \\
Due to the large scale deployment of Femtocells in enterprise/office environments~\cite{tahalani2014optimal,sathya2015joint,tahalani2013optimal} and many practical constraints (\emph{e.g.,} lack of space and power), operators will go for arbitrary deployment. Arbitrary deployment of Femtos will lead to coverage holes and an increased number of Femtos (increased OPEX and CAPEX). To address these issues, placement of Femtos needs to be optimal \cite{lokhandwala2014phantom, optimal}. Optimal placement of Femtos~\cite{tahalani2014optimal,sathya2014placement,sathya2020small} ensures good SINR and improves overall system capacity. Usually, due to physical constraints, operators may need to go for sub-optimal or arbitrary deployment. Consequently, the number of Femtos deployed is more than that in the optimal model to ensure no coverage holes inside the building. Approaches that deal with the deployment of single Femto is not scalable to enterprise buildings. Placement approaches do not consider the building model parameters (like wall thickness), dynamic power transmission, cost of human resources, and field testing for randomly placed Femtos. Femtos create coverage holes for UEs when the Femtos operate in closed access mode, and non-subscriber UEs are in close vicinity. \cite{lokhandwala2015phantom,sathya2013efficient,ramamurthy2015energy} propose to mitigate coverage holes by proper placement and power control mechanisms. \\

In HetNet systems, the major factor that affects the network throughput is the interference (between Femtos and between Macro and Femtos). There are two types of interference possible in the HetNet systems:\\
\begin{enumerate}
\item \textbf{Co-tier Interference:} Due to reuse of one usage of spectrum, interference from the neighboring small cells is called co-tier interference~\cite{sathya2016improving,sathya2016handover,sathya2020raptap}. For example, UE2 is getting served by the Femto BS (F2), but it is receiving interference from the neighboring Femto BSs (F1 \& F3). The traditional solution to avoid co-tier interference among BSs is Inter-Cell Interference Coordination (ICIC). In the ICIC scheme, all BSs cooperatively communicate using the X2 interface and allocate RBs efficiently to the cell edge users, but on the other hand, this increases the signaling messages.
\item \textbf{Cross-tier Interference:} The interference between Macro BS and small cell is called cross-tier interference ~\cite{sathya2015femto,sathya2016handover,sathya2016maximizing}. For example, UE1 is getting served by the Macro BS, but it is receiving interference from small cells (\emph{i.e.,} Femtocells F1, F3). The traditional solution to avoid cross-tier interference is enhanced ICIC (eICIC)~\cite{giluka2016handovers,giluka2016leveraging,sathya2020qos}. In the eICIC scheme, the interference between MBS and Femto BS (FBS) is avoided by muting some sub-frames (Almost Blank Sub-frame) in MBS during FBSs transmissions. In turn, it reduces the interference and increases the capacity in HetNet systems.\\
\end{enumerate}
 Though the spectrum efficiency and system capacity could increase due to spatial reuse of the same spectrum in LTE HetNets, SINR and network throughput may be affected by cross-tier inference MBS(s) and small cells and co-tier interference among small cells and obstacles inside buildings. The signal leaks at the buildings' edges/corners, which causes cross-tier interference and degrading the performance of $OUEs$ in the High Interference Zone (HIZone) around the building connected to one of the Macro BSs in LTE HetNet. To the best of our knowledge, none of the existing works addressed the cross-tier interference issue to HIZone UEs ($HIZUEs$) in a dynamic fashion based on their occupancy levels in the HIZone. In this work, we propose an active power control scheme which is employed at the Femtos to reduce cross-tier interference to $HIZUEs$ in the HIZone.\\

\item Scheduling or Radio Resource Allocation \\ \\
Since Femtos are deployed to offer high data rate services to indoor (paid) users in enterprise and residential buildings, each Femto is configured with a list of subscribers called Subscriber Group (SG) that can access them~\cite{ghosh2017novel}. The users not belonging to this list belong to the Non-Subscriber Group (NSG), and they are served by MBSs even when they are close to a Femto. This type of restricted access is called closed access. Femtos configured in open access do not distinguish between SG and NSG users, and hence, they may fail to ensure QoS for SG users, especially during peak traffic loads. 
%Hybrid access mechanism integrates the principles of both closed and open access mechanisms. It lets some NSG users connect with Hybrid Access Femtocells (HAFs) and share its radio resources and SG users. This mechanism provides a trade-off between maximizing overall HetNet capacity and maximizing the throughput of SG users. Figure~\ref{figoac} shows an LTE HetNet system comprising one Macrocell and three Femtocells: one OAF, one CAF, and one HAF. Here the OAF is serving both SG and NSG users located in its coverage area. In the CAF case, all of the NSG users located in the CAF's coverage area are forced to connect with the MBS. As a compromise, HAF serves two NSG users and SG users, and the rest of NSG users are filled by the MBS. 
%Femtos that are configured in open access mode does not distinguish between SG users and non-SG users, and hence, they may fail to ensure QoS for SG users, especially during peak traffic loads.
Telecom operators favor hybrid access Femtocells (HAFs) as they can provide QoS for SG users by giving them preferential access to radio resources over NSG users and also improve the capacity of LTE HetNet by serving nearby NSG users. Rewarding mechanisms have been proposed to popularize hybrid access mode for Femtos. Challenges here are optimal HAF deployment and efficient splitting of radio resources~\cite{New222, New333, New444, New555, New666} between SG and NSG users of HAFs in indoor environments. \\
%To the best of our knowledge, none of the existing works discussed the fair allocation of radio resources among SG and NSG users to the best of our ability. This work proposes a dynamic bandwidth allocation method that divides the available bandwidth between the SG and NSG users and efficient power control and appropriate resource allocation method allocating the radio resources between SG and NSG users. 

 \item Improving Data Rates in LTE Small Cells and D2D Communication \\ \\
 Though the deployment of Femtocells (Femtos) improves indoor data rates, the resulting LTE Heterogeneous Network may face a host of problems~\cite{sathya2016improving,kala2019statistical}. To number a few co-tier, cross-tier interference (frequent reuse in one LTE), and regular handovers due to short coverage areas of Femtos are some significant issues. Deployment of Femtos inside a building can lead to signal leakage at the buildings' edges/corners. It causes cross-tier interference and degrades outdoor UEs in the HIZone around the building area, connected to one of the Macro BS (MBSs) in the LTE HetNet. Arbitrary placement of Femtos can lead to high co-channel cross-tier interference among Femtos~\cite{rangisetti2020qos,sathyamodified} and MBSs coverage holes inside buildings. If Femtos are placed without power control, it leads to increased power consumption and high inter-cell interference in large scale deployments. Our goal was to address these problems by developing efficient LTE small cell architecture, optimal Femto placement, power control to ensure SINR threshold in Indoor environment and Device-to-Device (D2D) based communication were in free/idle Indoor UEs connected to Femto act like UE-relays (\emph{i.e.,} UE-like BS, forwarding downlink data plane traffic for some of the phone users connected to MBS.\\

\item Asymmetric Vs. Symmetric ED threshold on LAA and Wi-Fi \\ \\
The exponential increase in the number of mobile devices in use today has led to a commensurate increase in cellular and Wi-Fi infrastructure demands, thus requiring that both licensed (cellular) and unlicensed (Wi-Fi) spectrum be utilized as efficiently as possible. The industry's actively pursued solution is for cellular systems to use the unlicensed spectrum and the licensed spectrum, which would require fair coexistence with Wi-Fi in the unlicensed spectrum. As per the IEEE 802.11 standard, Wi-Fi uses an energy detection (ED) threshold of -62 dBm when LTE-LAA and LTE-U nodes are deployed close by LTE-LAA specification recommends that LTE-LAA detect Wi-Fi at -72 dBm. In our work~\cite{iqbal2017impact}, we evaluate the effect of this asymmetry in the ED threshold on the coexistence between two systems. We develop a coexistence simulator in ns-3 and vary both the Wi-Fi and LTE energy detection thresholds, and demonstrate that lowering the Wi-Fi ED threshold from -62 dBm improves performance Wi-Fi LTE-LAA. Prior work has mostly focused on determining the ED threshold that should be used by LTE-LAA/LTE-U. As far as we are aware, this is the first result that demonstrates that lowering the Wi-Fi ED threshold improves both systems' performance. The conclusion is that if Wi-Fi treats LTE-LAA/LTE-U as it would an overlapping Wi-Fi, coexistence performance improves compared to the current assumption that Wi-Fi treats LTE-LAA/LTE-U as noise.\\

\item Facilitating LAA/Wi-Fi Coexistence Using Machine Learning Approach \\ \\
The various aspects of the coexistence scenarios such deployments give rise to have been considered in a vast body of academic and industry research. However, there is very little data and analysis on how these coexisting networks will behave in practice. The question of "fair coexistence" between Wi-Fi and LAA has moved from a theoretical question to reality. The recent roll-out of LAA deployments provides an opportunity to collect data on these networks' operations and study coexistence issues on the ground. In our recent work~\cite{sathya2020measurement}, we examined the problems raised due to the coexistence of LAA and Wi-Fi in a real-time deployment scenario by various US carrier operators in downtown Chicago. Based on the nature of the traffic (e.g., data, video, live streaming, etc.) and the availability of the small cell (Femto/LAA) coverage, enabling of secondary unlicensed component carriers are observed. The observation concluded that due to a static channel allocation strategy followed by an  LAA  BS, a particular channel occupied for a longer time, thereby unlicensed Wi-Fi APs face resource crunch as they follow dynamic channel allocation strategy. As a solution, we predicted the best channel assignment to LAA BS by applying ML algorithms to the collected data so that Wi-Fi users are not affected.\\

\item Association Issues on LTE-U/Wi-Fi Coexistence \\ \\
In our previous  work~\cite{sathya2018association,sathya2018analysis},  we address the issue of association fairness when  Wi-Fi and  LTE  unlicensed  (LTE-U)  coexist on the same channel in the unlicensed  5  GHz band.  Since beacon transmission is the first step in starting the association process in  Wi-Fi,  we define association fairness as to how fair  LTE-U  is in allowing  Wi-Fi to start transmitting beacons on a  channel that it occupies with multiple duty cycles.  According to the LTE-U  specification,  if an LTE-U  base station determines that a  channel is vacant,  it can transmit for up to  20  ms and turn OFF for only 1 ms, resulting in a duty cycle 95\%. In an area with heavy spectrum usage,  there will be cases when a  Wi-Fi access point wishes to share the same channel,  as it does today with  Wi-Fi.  We study,  both theoretically and experimentally, the effect that such a sizeable LTE-U  duty cycle can have on the association process,  specifically  Wi-Fi beacon transmission and reception.  We demonstrated via an experimental set-up using NI USRPs so that a significant percentage of Wi-Fi beacons will neither be transmitted in a timely fashion nor be received at the LTE-U BS, thus making it difficult for LTE-U BS  to adapt its duty cycle in response to the Wi-Fi usage.
We proposed a novel Carrier Sense Adaptive Transmission (CSAT) algorithm~\cite{sathya2020wi,manas2018socio,kala2020cirno} to address the problem of Wi-Fi client association in a dense deployment scenario and enable a fair share of spectrum access.\\
 
 \item Optimal Scaling of LTE-U Duty cycle in LTE-U/Wi-Fi Coexistence \\ \\
 
 The application of Machine Learning (ML) techniques to complex engineering problems has proved to be an attractive and efficient solution. ML has been successfully applied to several practical tasks like image recognition, automating industrial operations, etc. The promise of ML techniques in solving non-linear problems influenced this work, which aims to apply known ML techniques and develop new ones for wireless spectrum sharing between Wi-Fi and LTE in the unlicensed spectrum. In this work~\cite{sathya2019auto,adam2019detection,sathya2020machine,dziedzic2020machine}, we focus on the LTE-U specification developed by the LTE-U Forum, which uses the duty-cycle approach for fair coexistence~\cite{sathya2018energy,kumar2019enhancing}.  The operator can scale the LTE-U duty cycle optimally if the exact numbers of Wi-Fi APs are known. In the literature, no work is proposed on identifying the precise number of APs by the LTE-U BS.  The specification suggests reducing the LTE-U BS's duty cycle when the number of co-channel Wi-Fi basic service sets (BSSs) increases from one to two or more. However, without decoding the Wi-Fi packets, detecting the number of Wi-Fi BSSs operating on the channel in real-time is challenging. This work demonstrates a novel ML-based approach that solves this problem using energy values observed during the LTE-U OFF duration. It is relatively straightforward to watch only the energy values during the LTE-U BS OFF time compared to decoding the entire Wi-Fi packet, which would require a full Wi-Fi receiver at the LTE-U base-station. We implement and validate the proposed ML-based approach by real-time experiments and demonstrate distinct patterns between the energy distributions between one and many Wi-Fi AP transmissions. The presented ML-based approach results in higher accuracy (close to 99\% in all cases) as compared to the existing auto-correlation (AC) and energy detection (ED) approaches.\\

\end{enumerate}

\section{Future NR-U Small Cell in 6 GHz}
 The current research (on Spectrum Sharing on 5 GHz) has opened many exciting possibilities to solve the research challenges for LTE/Wi-Fi coexistence in 6 GHz used as an unlicensed band~\cite{garg2019sla,kala2019odin}. In the following, we outline some of my future directions on the spectrum, sharing small cells.

\begin{figure}
\begin{center}
\includegraphics[width=0.5\textwidth]{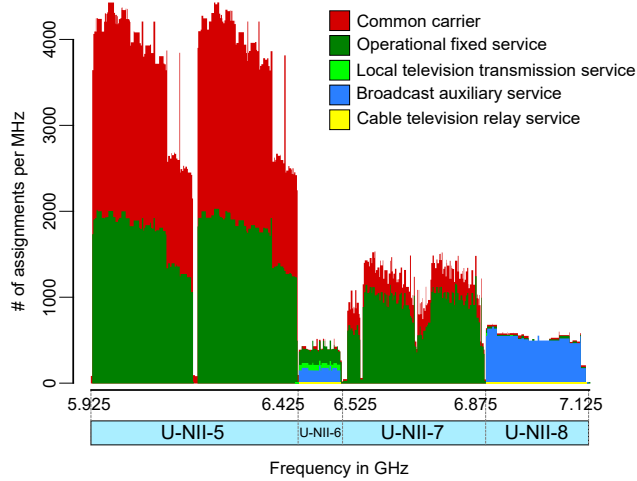}
%  \vspace{-0.5cm}
  \caption{Bandwidth Allocation on 6 GHz Spectrum \cite{FCC1}}
  \label{6ghz}
  \end{center}
\end{figure}

 \subsection{Fair Coexistence of NR-U and Wi-Fi in 6 GHz Spectrum~\cite{patriciello2020nr}:}
 
 Since the licensed spectrum is a limited and expensive resource, its optimal utilization may require spectrum sharing between multiple network operators/providers of different types. Increasingly licensed-unlicensed sharing is being contemplated to enhance network spectral efficiency beyond the more traditional unlicensed-unlicensed sharing. As the most common unlicensed incumbent, Wi-Fi is now broadly deployed in the unlicensed $5$ GHz band in North America, where approximately $500$ MHz of bandwidth is available. However, these $5$ GHz unlicensed bands are also seeing the increasing deployment of cellular services such as LTE-LAA. Recently, the Federal Communications Commission (FCC) sought to open up 1.2 GHz of additional spectrum for unlicensed operation in the 6 GHz~\cite{patriciello2020nr,coex} band through a Notice of Proposed Rule Making (NPRM) \cite{FCC1,WA,sathya2020standardization} as shown in Fig.~\ref{6ghz}.  Thus, this spectrum allocation for the unlicensed operation will only accelerate the need for other coexistence solutions among heterogeneous systems. Hence it is clear that regulatory authorities worldwide are paying close attention to the 6 GHz band as the next spectrum band that will continue to enhance unlicensed services across the world. However, it is also clear that this band, like the 5 GHz band, will see both Wi-Fi and cellular systems being deployed, and hence the coexistence issues played out in the 5 GHz band will repeat in this new frequency as well. In recognition of this, the two principal stakeholder standardization entities, IEEE and 3GPP, held a coexistence workshop in July 2019 \cite{coex} to discuss methods to address this before the next generation standards being specified. This section discusses the recent activities on FCC's 6 GHz NPRM and IEEE \& 3GPP efforts towards coexistence in the 6 GHz band.

\subsection{6 GHz Coexistence: Deployment Scenarios, and Channel Access}

Although several industry entities were not in favor of a re-evaluation, IEEE recommended that coexistence evaluations for NR-U should include 802.11ac (in 5 GHz), 802.11ax (in 6 GHz), and 802.11ad (in 60 GHz). For the sub-7 GHz bands, coexistence evaluations will be technology-neutral (e.g., channel access mechanism) and performed in random carrier frequencies in the 5 GHz band. These evaluations also necessitate devising suitable 11ac/ax coexistence topologies with a significant number of links below -72 dBm. 
\subsection{Future NR-U: Deployment Scenarios}
The NR-U work item recently approved by 3GPP supports the existing unlicensed 5 GHz band and the new unlicensed "\textit{greenfield}" 6 GHz band. Industry players such as Qualcomm expect that other unlicensed and shared spectrum bands, including mmWave, will be added to this list in future releases. Researchers will study the following deployment scenarios to investigate the functionalities needed beyond the operation specifications in an unlicensed spectrum. 
\begin{itemize}
\item \textit{Carrier aggregation} between licensed band NR (PCell) and NR-U (SCell): (a) NR-U SCell with both DL and UL. (b) NR-U SCell with DL-only.
\item \textit{Dual connectivity} between licensed band LTE (PCell) and NR-U (PSCell)
\item \textit{Stand-alone} NR-U
\item An NR cell with DL in the unlicensed band and UL in licensed band
\item \textit{Dual connectivity} between licensed band NR (PCell) and NR-U (PSCell)
\end{itemize}
The Legacy cellular operators oppose the NR-U stand-alone scenario and want 3GPP to drop it. They fear stiff competition from new players who can use NR-U stand-alone for limited cellular operation. NR-U is likely to be a more potent competitor to 802.11 than LAA as it will have a more flexible and efficient PHY/MAC marked by a shorter symbol duration, shorter HARQ Round Trip Time (RTT), etc. Further, NR-U can be deployed in every configuration where 802.11 is currently operational if both stand-alone and dual connection is approved. Also, unlike 802.11, NR-U will be capable of deploying the same PHY/MAC with flexible configurations across all current and future unlicensed bands.

\subsection{ML Based approaches to Solve Issues in Current and Future Spectrum Sharing:}

We have listed some of the interesting problems on LAA/Wi-Fi coexistence solved through an ML-based approach:

\begin{itemize}
    \item \textbf{Narrowband vs. Wideband LBT in 6 GHz:} The LBT mechanism is used by a device to avoid collisions by ensuring that no other transmissions are concurrently active in the channel. LTE-LAA follows CAT 4 LBT for most of its transmissions, while CAT 2 LBT is used for about 5\% of DL transmissions. NR-U is likely to adopt a mechanism similar to the LAA LBT. NR-U Release 16, like its predecessor, the NR Release 15, supports component carriers up to the maximum limit of 100 MHz bandwidth. Besides, it supports the aggregation of several inter and intraband component carriers. Multi-carrier LBT channel access as defined in 5 GHz is assumed \emph{i.e.,} the Type A LBT in 3GPP TS37.213, where each channel performs its independent LBT procedure. Consequently, there is bound to be high complexity when the operation bandwidth is wide. The alternative Type B LBT in 3GPP TS37.213 can reduce this complexity by performing a single LBT on multiple channels. The wideband LBT could simplify the implementation of wideband operation when it identifies that the channel is free of narrowband interference \emph{i.e.,} limiting the narrowband signal (20 MHz) certain sub-bands or by long/short term measurements and LBT bandwidth adaption.
Hence, wideband LBT is beneficial for systems operating with wide bandwidth as it simplifies LBT implementation.
    \item \textbf{Intelligent Selection of Unlicensed Channel by LAA BS:} While analyzing the collected data in real-time on LAA/Wi-Fi coexistence~\cite{sathya2020measurement}, we found that in an incredibly dense deployment scenario, multiple LAA operators contest for the unlicensed channel. However, numerous unlicensed channels are available, but choosing a particular channel and estimating the duration to occupy the track, with the vision of not affecting the Wi-Fi users and multiple LAA operators, is challenging. To solve this problem, we propose to use a Q-learning based ML solution so that a channel and its occupancy time decided intelligently based on the parameters such as interference from other operators, load on the channels, channel activities of Wi-Fi users, etc.
    \item \textbf{Intelligent Channel Selection by Wi-Fi AP in LTE/Wi-Fi Coexistence:} In a multi-AP setting, an AP selects the channel for the operation of the expected capacity of the existing links. The traditional way is to take the SINR-based capacity estimate into account. However, this capacity model may sometimes fail to represent the complex interactions between PHY and MAC layers due to the presence of LAA, as it makes the scenario heterogeneous. As a result, decisions regarding channel selection may be delayed or inaccurate. To solve this problem, we propose to use supervised learning as a tool to model the complex interactions between PHY and MAC layers based on factors such as power and PHY rate of a neighboring Wi-Fi link.
    \item \textbf{Efficient Radio Resource Allocation Using Reinforcement Learning} In the fifth-generation (5G) of mobile broadband band systems, Radio Resources Management (RRM) has reached unprecedented levels of complexity. To cope with the ever more sophisticated RRM functionalities and to make prompt decisions required in 5G, efficient radio resource scheduling will play a critical role in the RAN system. Depending on tasks' purposes, the scheduling process is divided into three steps: prioritization, resource number determination, and resource allocation. However, to support a diverse range of applications such as ultra-reliable low latency applications, IoT applications, V2X applications, AR/VR applications, massive multimedia applications, etc., the scheduling process's steps become more complex. As a solution, we propose to use deep reinforcement learning tools to get feedback about the traffic in real-time so that efficient scheduling decisions and optimal use of radio resources are made. 
\end{itemize}

\section{Conclusions}
Small Cells deployment in 4G and 5G play a crucial role in boosting the capacity by efficient reuse of the same spectrum. This paper listed some of the open challenges and issues in the current (in 5GHz) and future (in 6GHz) spectrum sharing. We will continue the research into coexistence by shifting focus onto the latest unlicensed band: 6 GHz. The newest Wi-Fi version, 802.11ax, will coexist with the latest cellular technology, 5G NR-U. Unlike LAA, 5G NR-U will transmit both uplink and downlink data in the unlicensed band. Unlike 802.11ac in 5 GHz, 802.11ax in 6 GHz will employ Orthogonal Frequency Division Multiple Access (OFDMA): these changes will create new coexistence scenarios.

\end{document}